\documentclass[10pt]{article}
\usepackage{amssymb}
\begin{document}

\makeatletter
\def\@maketitle{\newpage
 \null
 {\normalsize \tt \begin{flushright} 
  \begin{tabular}[t]{l} \@date  
  \end{tabular}
 \end{flushright}}
 \begin{center} 
 \vskip 2em
 {\LARGE \@title \par} \vskip 1.5em {\large \lineskip .5em
 \begin{tabular}[t]{c}\@author 
 \end{tabular}\par} 
 \end{center}
 \par
 \vskip 1.5em} 
\makeatother
\topmargin=-1cm
\oddsidemargin=1.5cm
\evensidemargin=-.0cm
\textwidth=15.5cm
\textheight=22cm
\setlength{\baselineskip}{16pt}
\title{The Holographic Fluid on the Sphere Dual to the Schwarzschild Black Hole}
\author{Ryuichi~{\sc Nakayama}\thanks{ nakayama@particle.sci.hokudai.ac.jp}
       \\[1cm]
{\small
    Division of Physics, Graduate School of Science,} \\
{\small
           Hokkaido University, Sapporo 060-0810, Japan}
}
\date{
EPHOU-11-006  \\
September  2011  
}
%
%
\maketitle

\begin{abstract} 
We consider deformation of the d+2 dimensional asymptotically flat 
Schwarzschild black hole spacetime with the induced metric on a d-sphere 
at $r=r_c$ held fixed. This is done without taking the near horizon limit. 
The deformation is determined so that the $\Lambda=0$ vacuum Einstein 
equation is satisfied and the metric is regular on the horizon.  
In this paper the velocity of a dual fluid $v^i$ is assumed to be a 
Killing field and small, and the deformed metric is obtained up to  
$O(v^2)$. At this order of hydrodynamic expansion the dual fluid is an 
ideal one. The structure of the metric is fairly different from 
the near horizon result of Bredberg and Strominger in arXiv:1106.3084.

\end{abstract}
\newpage
\setlength{\baselineskip}{18pt}

\newcommand {\beq}{\begin{equation}}

\newcommand {\eeq}{\end{equation}}
\newcommand {\beqa}{\begin{eqnarray}}
\newcommand {\eeqa} {\end{eqnarray}}
\newcommand{\bm}[1]{\mbox{\boldmath $#1$}}
\newcommand{\Sq}{D(X)}
\newcommand{\al}{2\pi \alpha'}
\newcommand{\RR}{{\mathsf R\hspace*{-0.9ex}%
  \rule{0.15ex}{1.5ex}\hspace*{0.9ex}}}
\section{Introduction}
\hspace{5mm}
In theories which include gravity, such as string thery, 
in appropriate background spacetime such as asymptotically Anti-de Sitter
space, the bulk theory has a dual description in terms of the field theory 
at the asymptotic boundary. This is known as AdS/CFT, or AdS/FT 
duality.\cite{Mardacena}\cite{Gubser}\cite{Witten} 

Recently, the bulk theory of gravity theory can be equally described in terms of 
fluid theory at the asymptotic boundary. 
This is called Fluid/Gravity duality.\cite{Wadia}\cite{HUBE} Such studies began in 
the context of 
Yang-Mills theory at finite temperature. By using this duality several 
quantities such as shear viscosity to entropy density ratio $\eta/s$ 
have been computed.\cite{Poli} The novelty of this approach is the 
weak/strong duality and various quantities in strong coupled theory can be 
computed by calculations in a weak (or classical) gavity theory. 
The holographic idea behind the Fluid/Gravity duality was originally presented 
in connection with the black hole horizon.\cite{PW}\cite{Dam}\cite{PT}\cite{TPM}

In a recent paper \cite{NStoEinstein} a remarkable relation between incompressible 
non-relativistic fluids in d+1 dimensions satisfying the Navier-Stokes equation and 
d+2 dimensional Ricci flat metrics was found. 
The argument of \cite{Bremdberg}\cite{NStoEinstein} starts from a Rindler space 
and at radial distance $r=r_c$ a cutoff flat hypersurface $\Sigma_c$ is considered.
They studied the effect of perturbations of extrinsic curvature of $\Sigma_c$ 
while keeping the induced metric on $\Sigma_c$ fixed.  It was found that if the 
Brown-York stress tensor\cite{BY} on $\Sigma_c$ is that of the incompressible 
Navier-Stokes fluid, a corresponding Ricci flat metric exists which is regular at the
Rindler horizon. Afterwards, systematic algorithm for reconstructing a solution of the d+2 
dimensional vacuum Einstein equations from a d+1 dimensional fluid to arbitrary order was 
presented.\cite{SkenderisTaylor} In \cite{SkenderisTaylor} it was also shown that the 
seed metric can be constructed by a constant boost in a hyperplane in the Rindler space. 
Then in \cite{Petrov} a connection between Petrov type I condition on 
solutions of Einstein equation and incompressible Navier-Stokes equation was 
elucidated.

Then a possible generalization of these works will be to take a curved manifold as
the cutoff hypersurface $\Sigma_c$. The simplest choice of $\Sigma_c$ will be $\RR \times S^d$ with
$\RR$ for time.  Unperturbed spacetime will be that of an asymptotically flat 
Schwarzschild black hole. We can choose 
as $\Sigma_c$ a sphere $S^d$ at radial distance $r=r_c$ from the origin. 

In \cite{Sphere} such construction of a deformed metric was done in the near horizon limit and 
expansion around it. However, duality between a Schwarzschild black hole and a fluid on a 
sphere is expected to exist even without taking the near horizon limit.
The purpose of the present paper is to study perturbations of Schwarzschild metric 
with the induced metric on a spherical $\Sigma_c$ kept fixed, without taking the near horizon limit.  
In sec. 2  we will deform the Schwarzschild metric by introducing terms which depend on 
the vector field $v^i$ and pressure $P$, while keeping the induced metric on $\Sigma_c$ 
fixed and still making the metric satisfy vacuum Einstein equation and be regular at 
the horizon. We will regard the vector field $v^i$ and pressure $P$ as small, and 
 being of order $O(\epsilon^1)$ and $O(\epsilon^2)$,
respectively. We will consider metric corrections up to $O(\epsilon^2)$. 
 The equations which determine the coefficient functions of the tensors are presented
and solved. 
Then from the above results we will find that the Brown-York stress tensor\cite{BY} on $\Sigma_c$  
takes the form of an incompressible ideal fluid.  
In sec. 3 we end with a discussion.

\section{Deformed metric}
\hspace{5mm}
In this section we will start with a Schwarzschild black hole metric in the 
ingoing Eddington-Finkelstein coordinates
\begin{eqnarray}
ds^2= -f(r) \, dt^2+2dt \, dr+r^2 \, d\Omega_d^2
\label{Schwarz}
\end{eqnarray}
and construct a new metric which contains a pressure $P$ 
and the velocity $v^i$ and coincides with the Schwarzschild metric on a 
hypersurface $r=r_c$.  Here 
\begin{eqnarray}
f(r)=1-\left(\frac{r_h}{r}\right)^{d-1} \equiv 1-\frac{M}{r^{d-1}}
\label{ef}
\end{eqnarray}
 and $r=r_h$ is the event horizon. 

We use the standard round metric $d\Omega_d^2=h_{ij} \, dx^i \, dx^j$ on S$^d$, given by 
$h_{ij}=\delta_{ij}+\frac{x^i \,  x^j}{1-x^2}$, $(i,j=1,2, \ldots, d)$. 
The inverse metric is given by $h^{ij}=\delta^{ij}-x^i \, x^j$. 
The index of $v^i$ is raised and lowered by these metrics. The norm 
of $v^i$ is given by $v^2=h_{ij} \, v^i \, v^j$. Throughout this paper 
$v^i$ is assumed to be a killing vector field of S$^d$. 
\begin{eqnarray}
\nabla^{(h)}_i \, v_j+\nabla^{(h)}_j \, v_i=0
\end{eqnarray}
Here $\nabla^{(h)}_i$ is the covariant derivative for $h_{ij}$. 
The norm of the velocity $v^2$ and pressure $P$ are constant. 
Furthermore these are 
assumed to be small and are of orders $v^i = O(\epsilon^1)$ and  
$P =O(\epsilon^2)$, 
where $\epsilon$ is a hydrodynamic expansion parameter.

\subsection{Pressure}
\hspace{5mm}
We first consider the part of the metric which contains the pressure $P$.  
In the calculation up to $O(\epsilon^2)$, the pressure and the velocity 
can be considered separately. 

We make the following ansatz.
\begin{eqnarray}
dS^2= \left(-f(r)+a_1(r)P\right)dt^2+2 \left(1+a_2(r)P\right)dtdr+\left(r^2 +a_3(r) \, P\right)\, d\Omega_d^2
\label{metricP}
\end{eqnarray}
On the hypersurface ($\Sigma_c$) at $r=r_c$ this metric must coincide with the 
induced metric at $r=r_c$, 
\begin{eqnarray}
ds^2|_{r_c}= -f(r_c) \, dt^2+r_c^2 \, h_{ij} \, dx^i dx^j
\end{eqnarray}
and this leads to the condition $a_1(r_c)=a_3(r_c)=0$. To determine 
the functions $a_1(r)$, $a_2(r)$, $a_3(r)$, we must solve the vacuum Einstein equation 
with vanishing cosmological constant $\Lambda=0$. By the Einstein equation, we obtain
\begin{eqnarray}
&& \frac{d-1}{r} \ (1-f) \,  a_2'+a_1''+\frac{d}{r} \, a_1'
-d \, f' \, \left(\frac{1}{2r^2} \, a_3 \right)'=0, \\
&& a_2'=0, \qquad \left(\frac{1}{2r^2} \, a_3 \right)''+ \frac{2}{r} \, \left(\frac{1}{2r^2} \, a_3 \right)'=0, \\
&& (d-1) \, a_1+r a_1'+rf \, a_2'+2(d-1) \, a_2-\frac{1}{2} \, (fa'_3)'+\frac{2-d}{2r} \, fa'_3 \nonumber \\
&& +(2-d)rf \,  \left(\frac{1}{2r^2} \, a_3 \right)'  =0
\end{eqnarray}
Here $a_1'=da_1/dr$, etc. Solving the above 
equations with the above mentioned boundary condition, we obtain
\begin{eqnarray}
&& a_1(r)=f(r)-f_c+\frac{1}{2} \, c_3 \, (d-1)r_cM\left(r^{-d}-r_c^{-d}\right), \label{a1} \\
&& a_3(r)=c_3 \, r \, (r-r_c), \qquad c_3 \equiv 2 \,\left( a_2 +\frac{M}{2r_c^{d-1}}\right)
\,\left(1+ \frac{(d-1)M}{2r_c^{d-1}}\right)^{-1} \label{a3}
\end{eqnarray}
Here we defined $f_c \equiv f(r_c)$. $a_2$ is constant and its value is left undetermined.  
The overall normalization of $a_1(r)$ is adjusted by using rescaling of $P$.
Notice that $a_1(r)$ and $a_3(r)$ are regular at the horizon $r=r_h$. 

Some Christoffel symbols on the hypersurface $\Sigma_c$ at $r=r_c$ are given by
\begin{eqnarray}
&& {\Gamma^r}_{ij} = -r_cf_c h_{ij}+2r_cf_c \, a_2 \, P \, h_{ij}-\frac{1}{2} \, f_ca'_3(r_c) \, P \, h_{ij}, \label{GammaP1} \\
&& {\Gamma^r}_{tt} =\frac{1}{2} f_cf_c'-\frac{1}{2}\, f_c \, a_1'(r_c) \, P- f_c \,f'_c \,a_2\,  P, \label{GammaP2} \\
&& {\Gamma^r}_{ti}=0 \label{GammaP3}
\end{eqnarray}
Here $f'_c \equiv f'(r_c)$. 

The extrinsic curvature $K_{ab}$ of the surface $\Sigma_c$ is obtained by 
\begin{eqnarray}
K_{ab}=\frac{-1}{\sqrt{g^{rr}}} \, {\Gamma^{r}}_{ab}  \label{K}    
\end{eqnarray}
and the Brown-York Stress Tensor\cite{BY} is given by 
\begin{eqnarray}
T_{ab}= \frac{1}{8\pi G} \, \left(  g^{cd}K_{cd} \, g_{ab}-K_{ab} \right) \label{T}
\end{eqnarray}

\subsection{Velocity field}
\hspace{5mm}
Let us now consider deformation of the metric up to $O(\epsilon^1)$. 
We make the following ansatz.
\begin{eqnarray}
ds^2= -f(r)\, dt^2+2dt \, dr +2A_1(r) \,  v_i \,  dx^idr
+2A_2(r) \, v_i \, dx^idt+h_{ij} \, dx^idx^j
\end{eqnarray}
On $\Sigma_c$ ($r=r_c$), we must set $A_2(r_c)=0$. We also choose a gauge\cite{SkenderisTaylor}
\begin{eqnarray}
g_{rr}=\partial_r g_{ri}=\partial_r g_{rt}=0
\end{eqnarray}
and so $A_1$ is a constant. 

The function $A_1(r)$ is determined by solving vacuum Einstein equations with $\Lambda=0$. The condition is 
\begin{eqnarray}
A''_2+\frac{d-2}{r} \, A'_2-\frac{2(d-1)}{r^2} \, A_2 =-\frac{2}{r} \, f' \, A_1
\end{eqnarray}
One can show that if $A_1 \neq 0$, a general solution $A_2(r)$ always has a logarithmic 
singularity. To avoid this one must set 
\begin{eqnarray}
A_1=0
\end{eqnarray}
Then the solution which satisfy the condition at $\Sigma_c$ is given by 
\begin{eqnarray}
A_2(r)=C \, r^2 \, \left(1-\left(\frac{r_c}{r}\right)^{d+1}\right)
\label{A2}
\end{eqnarray}
Here $C$ is a constant. 

We next consider the 2nd order corrections. With the above result we expect 
that solution that contains 
$v^i$ also exists in higher orders.  We make the following ansatz.
\begin{eqnarray}
ds^2 &=& \{-f(r)+A_3(r) \, v^2 \} \, dt^2
+ 2\{1+A_4(r) \, v^2 \} \, dt dr \nonumber  \\
&&+ 2\, A_2(r) \, v_i    \, dtdx^i \nonumber \\
&& +\{r^2+ A_5 (r) \, v^2 \} \, h_{ij} dx^idx^j +A_6(r) \, v_i \, v_j \, dx^i dx^j  
\label{metric2}
\end{eqnarray}
On the hypersurface 
$\Sigma_c$, we must impose $A_3=A_5=A_6=0$. $A_4$ is expected to be a constant,
but we do not assume this and later show that this is indeed the case. 
By using the $\Lambda=0$ vacuum Einstein equation a set of equations 
for these functions are derived. 

\begin{eqnarray}
d \, \{\frac{1}{2}A_5'-\frac{1}{r} \, A_5 \}'+\{\frac{1}{2}A_6'-\frac{1}{r} \, A_6 \}'-d \, r \, A'_4=0, \label{E1}
\end{eqnarray}
\begin{eqnarray}
&& \{A_4f'+A_3'-\frac{1}{r^2} \, A_2A_2'\}'+\frac{d}{r} \, 
\{A_4f'+A_3'-\frac{1}{r^2}A_2A_2'\} \nonumber \\
&& -f'\{ d(\frac{1}{2r^2}A_5'-\frac{1}{r^3}A_5)+(\frac{1}{2r^2}A_6'-\frac{1}{r^3}A_6) \} 
= 0, \label{E2}
\end{eqnarray}
\begin{eqnarray}
&&-\{fA_3'+f'A_3+2ff'A_4\}'-\frac{d}{r}\{fA_3'+f'A_3+2ff'A_4\} \nonumber \\
&&+ff'\{d(\frac{1}{2r^2}A_5'-\frac{1}{r^3}A_5)+(\frac{1}{2r^2}A_6'
-\frac{1}{r^3}A_6) \} +f'(A_3'+f'A_4) \nonumber \\
&&+\frac{1}{r^2} f''A_2^2+\frac{d-2}{r^3}f'A_2^2+\frac{2(d-1)}{r^4} A_2^2
+\frac{f}{r^2}(A_2')^2 \nonumber \\
&&= 0, \label{E3}
\end{eqnarray}
\begin{eqnarray}
\frac{2d}{r^2}A_6-\frac{1}{2}(fA_6')'-\frac{df}{2r}A_6'+\frac{2f}{r}A_6'-\frac{2f}{r^2}A_6=
\frac{1}{2}(A_2')^2-\frac{2}{r}A_2A_2'+\frac{2}{r^2}(A_2)^2,
\label{E4}
\end{eqnarray}
\begin{eqnarray}
&&-\frac{2}{r^2}A_6-drf\{\frac{1}{2r^2}A_5'-\frac{1}{r^3}A_5\}-rf \{\frac{1}{2r^2}A_6'-\frac{1}{r^3}A_6\} \nonumber \\
&& +\frac{4-d}{2r}fA_5'-\frac{2f}{r^2}A_5+\frac{2-d}{r^2}(A_2)^2+(d-2)A_3+2(d-2)fA_4-r f A_4' \nonumber \\
&&+ \{ -\frac{f}{2}A_5'-\frac{1}{r}(A_2)^2+rA_3+2rfA_4\}'
=0  \label{E5}
\end{eqnarray}

First, eq (\ref{E1}) can be once integrated to the form. 
\begin{eqnarray}
d \, A_4= d \left( \frac{1}{2r}A_5'-\frac{1}{2r^2}A_5\right)+
\left( \frac{1}{2r}A_6'-\frac{1}{2r^2}A_6\right)+c_4 \label{A4A5A6}
\end{eqnarray}
Here $c_4$ is a constant. 

Secondly, (\ref{E2}) $\times f$+(\ref{E3}) yields the equation.
\begin{eqnarray}
f' \, A_4'= \left\{\frac{-1}{r^2} \, A_2''+\frac{2-d}{r^3} \, A_2'+\frac{2(d-1)}{r^4} \, A_2 \right\} \, A_2
\end{eqnarray}
The rhs of this equation can be shown to vanish by using (\ref{A2}) and 
$A_4$ is indeed constant as expected.  Then by using (\ref{A4A5A6}) and the boundary conditions at $\Sigma_c$ 
the following equation is derived.
\begin{eqnarray}
d \, A_5+A_6= 2(d \, A_4-c_4) \, r \, (r-r_c)
\label{A5A6}
\end{eqnarray}

Thirdly, we will determine $A_6$ by solving (\ref{E4}). By using (\ref{A2}) the 
rhs of (\ref{E4}) can be simplified and we have
\begin{eqnarray}
\frac{2d}{r^2}A_6-\frac{1}{2}(fA_6')'-\frac{df}{2r}A_6'+\frac{2f}{r}A_6'-\frac{2f}{r^2}A_6=\frac{1}{2} \, C^2 (d+1)^2 \, r_c^{2d+2} \, r^{-2d}
\label{E4p}
\end{eqnarray}
The general solution is given by a sum of a special solution $A_6^{(0)}$ and a general solution $A_6^{(1)}$
to the homogeneous equation. A special solution is given by 
\begin{eqnarray}
A_6^{(0)}(r)= \frac{C^2}{M} \, r^{2d+2}_c \, r^{1-d}
\end{eqnarray}
On the other hand a general solution to the homogeneous equation is rather 
difficult to obtain. Changing variables from $r$ to $f$ using (\ref{ef}) we 
obtain the homogeneous equation.
\begin{eqnarray}
(d-1)^2 \, f(1-f)^2 \, \frac{d^2A_6^{(1)}}{df^2}- (d-1)(1-f)\{1-d+(d+3)f\} \, 
\frac{dA_6^{(1)}}{df}-4(d-f) \, A_6^{(1)}=0
\end{eqnarray}
This is a Heun's equation with singularities $f=0,1,1,\infty$. 
The characteristic exponents for the regular singularity $f=0$ (horizon) are $\lambda=0, \, 0$ and so one solution is regular at $f=0$ 
and another has a $\log f$ singularity in general. Let us denote this regular 
solution by $w(r)$. Then $A_6$, which is regular at the horizon and vanish at $\Sigma_c$, is 
given by 
\begin{eqnarray}
A_6(r)= \frac{C^2}{M} \, r_c^{d+3} \, \left\{ \left(\frac{r_c}{r}\right)^{d-1}-
\frac{w(r)}{w(r_c)} \right\} 
\label{A6}
\end{eqnarray}
Then $A_5$ is obtained by using (\ref{A5A6}). 
An explicit form of $w(r)$ is not obtained yet.

Finally, $A_3$ and constants $A_4$, $c_4$ can be obtained by solving (\ref{E2}) and 
(\ref{E5}) 
with all the previous results substituted. 
Eq (\ref{E5}) has the form $rA'_3+(d-1) \, A_3 =j(r)$, where $j(r)$ is a 
known function. The equation obtained by substituting a derivative of this into (\ref{E2}) 
is satisfied, only if 
\begin{eqnarray}
A_4=c_4=0
\end{eqnarray}
With the relation (\ref{A5A6}) this means that 
\begin{eqnarray}
A_5(r)= \frac{-1}{d} \, A_6(r)
\label{A5}
\end{eqnarray}
 Then the above equation for $A_3$ is solved to give
\begin{eqnarray}
A_3(r)= C^2 \, r^{1-d} \, (r^d-r^d_c)+ C^2 \, \frac{d-1}{2d} \, r_c^{2d+2} \, r^{1-d} \, 
(r^{-d-1}-r_c^{-d-1}) \label{A3}
\end{eqnarray}

Now we will make some observation. 
The final metric is the combination of (\ref{metricP}) and (\ref{metric2}). 
\begin{eqnarray}
ds^2 &=& \{-f(r)+A_3(r) \, v^2 +a_1(r) \, P\} \, dt^2\nonumber  \\
&&+ 2\, A_2(r) \, v_i    \, dtdx^i+ 2\{1 +a_2 \, P\} \, dt dr  \nonumber \\
&& +\{r^2+ A_5 (r) \, v^2 +a_3(r) \, P\} \, h_{ij} dx^idx^j +A_6(r) \, v_i \, v_j \, dx^i dx^j  
\label{metric3}
\end{eqnarray}
The coefficient functions are given in (\ref{a1}), (\ref{a3}), (\ref{A2}), (\ref{A6}), 
(\ref{A5}) and (\ref{A3}). These functions contain two constant parameters. 
One of the constants, $C$, can be set to 1 by rescaling $v^i$. 
However, the other constant $a_2$ is undetermined. This means that the deformation of the 
metric is not unique.  

Some Christoffel symbols relevant to the extrinsic curvature on $\Sigma_c$ are given by
\begin{eqnarray}
{\Gamma^r}_{ti} &=& \frac{-f_c}{2} \, A_2'(r_c) \, v_i, \nonumber \\
{\Gamma^r}_{tt} &=& \frac{1}{2} \, f_c \, f'_c- \frac{1}{2} \, f_c \, A_3'(r_c) \, v^2-\frac{1}{2}\, f_c \, a_1'(r_c) \, P- f_c \,f'_c \,a_2\,  P, \nonumber \\
{\Gamma^r}_{ij} &=& -r_c f_c \, h_{ij}-\frac{f_c}{2} \, A_5'(r_c)  \, v^2 h_{ij} -\frac{f_c}{2} \, A_6'(r_c) \, v_i \, v_j  \nonumber \\
&& +2r_cf_c \, a_2 \, P \, h_{ij}-\frac{1}{2} \, f_ca'_3(r_c) \, P \, h_{ij}
\label{Christoffel}
\end{eqnarray}
Here the contribution of the Pressure $P$ (\ref{GammaP1})-(\ref{GammaP3}) is added. 
These are evaluated at $r=r_c$, and $A_2(r_c)=A_3(r_c)=0$ is used. 
The above result (\ref{Christoffel}) is not changed, even if the condition that $v^i$ is a Killing field is removed. 
This will determine $K_{ab}$ (\ref{K}) and stress tensor $T_{ab}$ (\ref{T}). 
The fluid described by the stress tensor constructed from these $\Gamma$'s is 
that of an ideal fluid. Because $v^i$ is a killing field, the fluid is incompressible ($\nabla^{(h)}_iv^i=0$).  
The hydrodynamic equation coming from the conservation 
equation of this stress tensor will be  the Euler equation.

\section{Discussion}
\hspace{5mm}
In this paper by starting from a Schwarzschild black hole solution in d+2 dimensions 
we studied the purturbation of the extrinsic curvature of a cutoff surface $\Sigma_c$, 
which is S$^d$ located at $r=r_c$, with keeping the induced metric on $\Sigma_c$ fixed. 
We imposed a condition that the deformed metric is still a solution to the $\Lambda=0$ 
vacuum Einstein equation and is regular at the horizon. The near horizon limit is not 
taken. The structure of the metric obtained is fairly different from the one in 
\cite{Sphere}. For example, terms of the forms $v_i \, dr \, dx^i$ and $\chi_i \, dx^i dr$, etc,  
do not exist in (\ref{metric3}). 
This situation is quite different from the flat $\Sigma_c$ case.\cite{NStoEinstein}
\cite{SkenderisTaylor} It is not clear how the near horizon limit of (\ref{metric3})
will coincide with the result of \cite{Sphere}. The deformed metric (\ref{metric3}) also contains an
undetermined constant parameter $a_2$. It is not clear whether this constant can be 
fixed by some further requirement. 

We considered only the case of $v^i$ being a Killing field and $v^2$ being constant, 
and studied the structure of the extrinsic curvature up to $O(v^2)$. We found that 
the Brown-York stress tensor takes the form of that for a incompressible perfect fluid. 
The corresponding hydrodynamic equation will coincide with the Euler equation. 

Now by changing the vector $v^i$ from a killing field to more general one, 
and introducing derivative terms of $v^i$ and $P$ into the metric 
in such a way that vacuum Einstein equation is satisfied, 
shear may appear in the stress tensor. Then the transport coefficients such as the 
ratio $\eta/s$ will be obtained from the coefficient of the shear. 
In \cite{Sphere} an obstruction to a pure Dirichlet boundary condition on $\Sigma_c$ was noticed 
at higher orders. It will be interesting to investigate whether a similar obstruction would occur 
even if the near horizon limit is not taken. 
We hope to report on this in the future.

\end{document}